# *Exceptional points in transistor-metamaterial inspired transmission lines*

David E. Fernandes[1], Sylvain Lannebère[2], Tiago A. Morgado[2], Mário G. Silveirinha[1]

[1] *University of Lisbon, Instituto Superior Técnico and Instituto de Telecomunicações, Avenida Rovisco Pais, 1, 1049-001 Lisboa, Portugal*

[2]*Instituto de Telecomunicações and Department of Electrical Engineering, University of Coimbra, 3030-290 Coimbra, Portugal*

*E-mail: dfernandes@co.it.pt, sylvain.lannebere@co.it.pt, tiago.morgado@co.it.pt, mario.silveirinha@co.it.pt*

**Abstract**

Motivated by our recent findings in [Phys. Rev. Lett. 128, 013902, 2022], which introduces a new class of electromagnetic bulk materials whose response is similar to conventional semiconductor transistors, here we propose a one-dimensional (1D) version of such a material based on transmission lines coupled with FET isolators. We demonstrate that the response of this 1D system is nonreciprocal and non-Hermitian, analogous to the idealized transistor-metamaterial, and is also characterized by a broken time-reversal symmetry. We analyze the wave propagation in the system and find that the interaction between the eigenmodes can either lead to gain or loss depending on the propagation distance. Furthermore, it is also shown that the system may be operated at an exceptional point, wherein the response of the structure is singular, and the power gain is maximized. Finally, we demonstrate that the exceptional point coincides with the point of operation of typical microwave amplifiers, such as the distributed amplifier.

# I. Introduction

Hermiticity is a feature of physical systems related to energy conservation. In a Hermitian system the dynamics of the state vector is conservative and governed by unitary (linear) operations [1]. Consequently, the operator or matrix that relates the initial and final states of such system has real-valued spectrum and the associated eigenvectors form an orthogonal basis [1, 2]. In the case of electrodynamics, a Hermitian response requires that the constitutive parameters of the materials satisfy $\bar{\bar{\varepsilon}} = \bar{\bar{\varepsilon}}^{\dagger}$ and $\bar{\bar{\mu}} = \bar{\bar{\mu}}^{\dagger}$, with the superscript standing for the conjugate transpose, commonly referred to as the Hermitian conjugate. As is well-known, non-Hermitian responses ubiquitously appear in many bosonic systems such as nonequilibrium open systems [3-6] or in correlated electron systems [7-9]. Indeed, dissipation mechanisms are always present in open systems [6]. Non-Hermitian physics can often provide unique opportunities to precisely control the wave propagation in photonic platforms [5, 10], particularly to obtain nonreciprocal effects. Electromagnetic nonreciprocity can be obtained based on non-Hermitian platforms using Parity-Time ($\mathcal{P} \cdot \mathcal{T}$) symmetric systems [11-13], active systems [14-16] or systems with optical gain [17].

Furthermore, in [18] we recently proposed a mechanism to obtain a strong nonreciprocal and non-Hermitian response based on the interaction between a static electric bias and material nonlinearities, analogous to the response of standard MOSFETs (Metal-Oxide-Semiconductor Field-Effect Transistors). In Refs. [18, 19] it was demonstrated that the response of this MOSFET metamaterial (MOSFET-MTM) can either exhibit optical gain or dissipation depending on the relative phase of the electric field components inside the material. Such metamaterial may be useful to obtain robust electromagnetic isolation and amplification.

In this work we investigate a possible 1D implementation of the MOSFET-MTM based on transmission lines coupled with FET isolators. We theoretically show that the response of the structure is nonreciprocal, non-Hermitian and is characterized by a broken time-reversal symmetry. We demonstrate that the interaction between the system eigenmodes can enable regimes with gain and loss, depending on the propagation distance.

Another important feature of non-Hermitian systems is that their eigenstates are generally non-orthogonal and that their energy spectrum is complex-valued. Hence these systems may be characterized by peculiar features, with one of the most remarkable being the presence of exceptional points [20-22]. At an exceptional point, the non-Hermitian operator or matrix becomes defective, leading to a singularity in which two or more eigenvalues (and the associated eigenvectors) coalesce and become degenerate [20, 22]. Exceptional points have recently been on the spotlight in optics for the unique opportunities that they may provide in sensing [23-28]. Exceptional points have also been studied in the context of chiral transport and topological phenomena [29-35], unidirectional invisibility [36-39], formation of bulk Fermi arcs [40] and unusual quantum criticality [41-47]. It is important to mention that degeneracies in the eigenvalues may also be observed in Hermitian systems [21, 48], however the corresponding eigenvectors can still form an orthogonal basis. Here, based on the MOSFET metamaterial concept, we put forward a solution to operate two asymmetrically coupled transmission lines at an exceptional point. We show that at the exceptional point the response becomes singular and the system behaves as an amplifier whose gain increases with the square of the number of transistors, similar to the well-known distributed amplifier [49-52].

The article is organized as follows. In Sect. II we briefly describe the main characteristics and wave propagation properties of the MOSFET-MTM. In Sect. III we propose a possible implementation of the MOSFET-MTM based on capacitively-coupled transmission lines. We analyze the wave propagation within the structure, demonstrating the feasibility of operating it at an exceptional point. The operation at the degeneracy point is marked by a substantial power gain. Finally, in Sect. IV the conclusions are drawn. The time dependence of the fields is assumed of the type $e^{-i\omega t}$, with $\omega$ being the oscillation frequency.

## II. MOSFET METAMATERIAL

The MOSFET metamaterial (MOSFET-MTM) is an idealized material that exploits the interaction between an electric static bias and material nonlinearities, drawing inspiration from the response of standard MOSFETs [18]. In the MOSFET-MTM the relation between the polarization vector **P** and the electric field **E** is $\mathbf{P} = \varepsilon_0 \overline{\overline{\chi}}(\mathbf{E}) \cdot \mathbf{E}$, with the susceptibility tensor given by:

$$\overline{\overline{\chi}}(\mathbf{E}) = \begin{pmatrix} \chi_{xx}(E_z) & 0 & 0 \\ 0 & \chi_{yy} & 0 \\ 0 & 0 & \chi_{zz} \end{pmatrix} \quad (1)$$

The susceptibility tensor is a nonlinear function of the electric field as the component $\chi_{xx}(E_z)$ depends on the electric field along the $z$-direction $E_z$. It was shown in Ref. [18] that when this material is biased by a static electric field $\mathbf{E}_0$ in the $xoz$ plane ($\mathbf{E}_0 = E_{x0}\hat{\mathbf{x}} + E_{z0}\hat{\mathbf{z}}$), the effective linearized response of the material, i.e. the response of the material for small variations of the electric field around the biasing point $\delta \mathbf{P} = \varepsilon_0 \overline{\chi}_{\text{lin}}(\mathbf{E}) \cdot \delta \mathbf{E}$, with $\mathbf{E} = \mathbf{E}_0 + \delta \mathbf{E}$ and $\delta \mathbf{E} \ll \mathbf{E}_0$, may be described by the following linearized susceptibility tensor:

$$\overline{\chi}_{\text{lin}} = \begin{pmatrix} \chi_{xx}(E_{z0}) & 0 & \partial_{E_z}\chi_{xx}(E_{z0})E_{x0} \\ 0 & \chi_{yy} & 0 \\ 0 & 0 & \chi_{zz} \end{pmatrix}. \qquad (2)$$

As the displacement vector $\mathbf{D}$ relates the fields as $\mathbf{D} = \varepsilon_0 \mathbf{E} + \mathbf{P}$, it follows that the relative effective linearized permittivity of the MOSFET-MTM may be written as [18]:

$$\overline{\varepsilon} = \mathbf{1} + \overline{\chi}_{\text{lin}} = \begin{pmatrix} \varepsilon_{xx} & 0 & \varepsilon_{xz} \\ 0 & \varepsilon_{yy} & 0 \\ 0 & 0 & \varepsilon_{zz} \end{pmatrix}. \qquad (3)$$

As $\overline{\varepsilon} \neq \overline{\varepsilon}^\dagger$, the response of MOSFET-MTM is non-Hermitian. Furthermore, since $\overline{\varepsilon} \neq \overline{\varepsilon}^T$ the material is also nonreciprocal.

To analyze the consequences of this type of metamaterial response, let us consider the plane wave propagation problem. Since we are only considering the linearized response of the MOSFET-MTM, for notational simplicity we replace $\delta \mathbf{E} \to \mathbf{E}$. The fields in the material can be determined from the Maxwell equations:

$$\nabla \times \mathbf{E} = i\omega\mu_0 \mathbf{H}, \qquad (4a)$$

$$\nabla \times \mathbf{H} = -i\omega\varepsilon_0 \overline{\varepsilon} \cdot \mathbf{E}. \qquad (4b)$$

For propagation along the $y$-direction ($\nabla = \partial_y \hat{\mathbf{y}}$), Eqs. (4a-b) may be rewritten in an expanded form as:

$$\begin{aligned} i\partial_y E_x &= \omega\mu_0 H_z \\ i\partial_y E_z &= -\omega\mu_0 H_x \end{aligned}, \qquad (5a)$$

$$\begin{aligned} i\partial_y H_x &= -\omega\varepsilon_0 \varepsilon_{zz} E_z \\ i\partial_y H_z &= \omega\varepsilon_0 (\varepsilon_{xx} E_x + \varepsilon_{xz} E_z) \end{aligned}. \qquad (5b)$$

By defining a state vector $\mathbf{f} = (E_x, E_z, -H_z, H_x)^T$, the Maxwell equations can be written in a compact matrix form as $i\partial_y \mathbf{f} = \mathbf{M} \cdot \mathbf{f}$, with $\mathbf{M}$ equal to:

$$\mathbf{M} = -\omega \begin{pmatrix} 0 & 0 & \mu_0 & 0 \\ 0 & 0 & 0 & \mu_0 \\ \varepsilon_0 \varepsilon_{xx} & \varepsilon_0 \varepsilon_{xz} & 0 & 0 \\ 0 & \varepsilon_0 \varepsilon_{zz} & 0 & 0 \end{pmatrix}. \qquad (6)$$

Assuming $\mathbf{k} = k_y \hat{\mathbf{y}}$, so that $\partial_y = ik_y$, the eigenmodes in the bulk material can be determined from the nontrivial solutions of the characteristic equation $\det(\mathbf{M} + k_y \mathbf{1}_{4\times 4}) = \left(k_y^2 - \left(\frac{\omega}{c}\right)^2 \varepsilon_{xx}\right)\left(k_y^2 - \left(\frac{\omega}{c}\right)^2 \varepsilon_{zz}\right) = 0$. There are two eigenmodes (two solutions on $k_y^2$) which can be identified as ordinary and extraordinary waves [18]. Due to the non-Hermitian response, the corresponding eigenvectors are not orthogonal. As described in [18], such a property can give rise to unique wave phenomena, such as regimes with optical gain or dissipation depending on the relative phase of the (position dependent) electric field components in the material. In particular, the metamaterial can serve as a building block to realize optical isolators and other nonreciprocal and active devices [18].

## III. TRANSMISSION-LINE MODEL

In this section we suggest a straightforward approach to implement MOSFET-MTM devices in the microwave range based on transmission line (1D) technology and FETs. Furthermore, we study the unique wave phenomena that arise from the MOSFET-MTM response.

The idea is to exploit a parallelism between the Maxwell's equations in a MOSFET-MTM (Eqs. 5a-b) and the transmission line equations [52]. In a coupled transmission line configuration, such as the one shown in Fig. 1a, the wave equations for propagation along the $y$-direction read:

$$\partial_y V_1 = i\omega L_{11} I_1, \tag{7a}$$

$$\partial_y I_1 = i\omega (C_{11} V_1 + C_{12} V_2), \tag{7b}$$

$$\partial_y V_2 = i\omega L_{22} I_2, \tag{7c}$$

$$\partial_y I_2 = i\omega (C_{21} V_1 + C_{22} V_2). \tag{7d}$$

Here, $C_{21}, C_{12}$ represent the mutual capacitances per unit of length (p.u.l.), which describe the coupling between the two lines, and $C_{11}, C_{22}$ the capacitances p.u.l. of each line alone. The inductances p.u.l. of each line are $L_{11}, L_{22}$. We neglect the inductive coupling between the lines, so that $L_{12} = L_{21} = 0$, as well as dissipation effects.

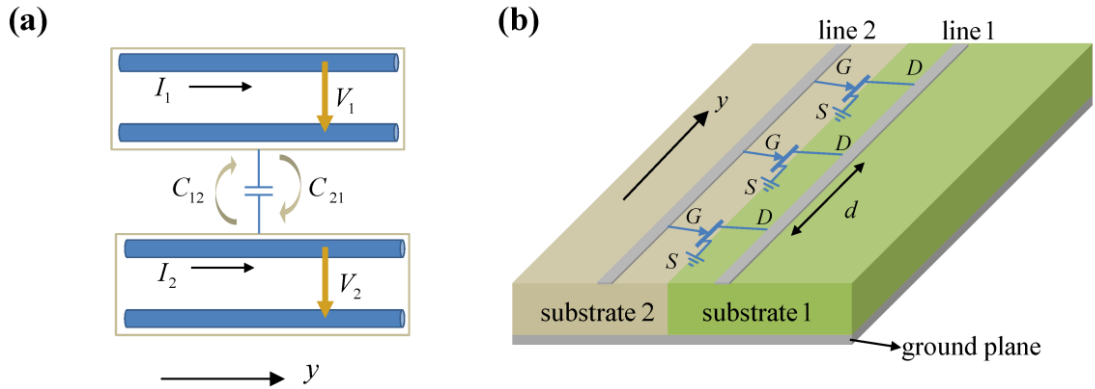

**Fig. 1.** (a) Two sets of transmission lines are capacitively coupled through mutual capacitances $C_{12}$ and $C_{21}$. (b) Transmission-line implementation of the MOSFET-MTM: two microstrip transmission lines oriented along the $y$-direction are periodically loaded with FET isolators (with spatial period $d$). We use a common-source configuration wherein the source is connected to the ground plane, the drain is connected to line 1 and the gate is connected to line 2.

It is easy to check that the electromagnetic fields and the material parameters can be mapped into equivalent transmission line voltages/currents and inductances/capacitances, respectively, by using the following transformation rules:

$$E_x \to V_1, \qquad E_z \to V_2$$
$$-H_z \to I_1, \qquad H_x \to I_2 \qquad (8a)$$

$$\mu_0 \to L_{11}, L_{22}$$
$$\varepsilon_0 \varepsilon_{xx} \to C_{11}, \quad \varepsilon_0 \varepsilon_{zz} \to C_{22}, \quad \varepsilon_0 \varepsilon_{xz} \to C_{12}, \quad \varepsilon_0 \varepsilon_{zx} \to C_{21} \qquad (8b)$$

Hence, in order to mimic the response of the MOSFET-MTM, the coupled lines are required to have $C_{12} \neq 0$ and $C_{21} = 0$. This type of asymmetric coupling causes the response of line 1 to be influenced by the waves propagating in line 2, whereas line 2 is completely isolated from line 1. One way to obtain this type of asymmetrically coupled response is by considering two microstrip lines periodically loaded with FET isolators, as illustrated in Fig. 1b. The transistors are 3 port devices, consisting of drain, gate and source. We adopt a common-source configuration, so that the source is connected to the ground plane of the microstrip lines. The gate is connected to line 2 (gate line) and the drain is connected to the line 1 (drain line). The response of the FET is modelled through an admittance matrix $\mathbf{Y}_F = \begin{pmatrix} Y_{11} & Y_{12} \\ Y_{21} & Y_{22} \end{pmatrix}$ that relates the currents and voltages at the two ports. The admittance matrix is determined from the small signal equivalent circuit of the FET, which results from a linearization of the response around an operation point [52], similar to what is done in the MOSFET-MTM. In Fig. 2 we depict the small signal equivalent circuit of the FETs under a common-source configuration [52].

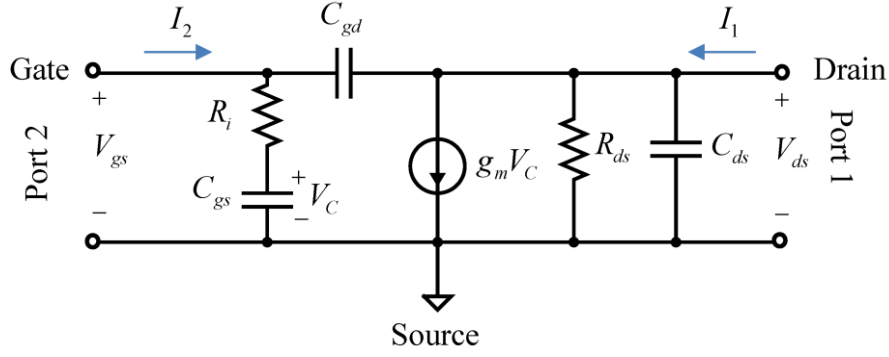

**Fig. 2.** Small-signal equivalent circuit for a microwave FET in the common-source configuration: $C_{gd}$ is the gate-to-drain capacitance, $C_{gs}$ is gate-to-source capacitance, $C_{ds}$ is drain-to-source capacitance, $g_m$ is the transconductance of the FET, $R_i$ is the series gate resistance and $R_{ds}$ is the drain-to-source resistance. We take port 1 as the drain of the FET and port 2 as the gate.

A straightforward analysis shows that the admittance matrix of the circuit depicted in Fig. 2 is given by:

$$\mathbf{Y}_F = \begin{pmatrix} \dfrac{1}{R_{ds}} - i\omega(C_{ds}+C_{gd}) & \dfrac{g_m}{(-i\omega C_{gs} R_i + 1)} + i\omega C_{gd} \\ i\omega C_{gd} & \left(R_i - \dfrac{1}{i\omega C_{gs}}\right)^{-1} - i\omega C_{gd} \end{pmatrix}. \tag{9}$$

where $C_{gd}$, $C_{gs}$ and $C_{ds}$ are the gate-to-drain, gate-to-source and drain-to-source capacitances, respectively. Moreover, $g_m$ is the transconductance of the FET, $R_i$ is the series gate resistance and $R_{ds}$ is the drain-to-source resistance. The gate-to-drain capacitance $C_{gd}$ is usually small and often can be ignored [52]. In the following, we neglect $C_{gd}$ except when explicitly stated otherwise. In that case, $Y_{F,21} \approx 0$ so that the gate becomes effectively isolated from the drain, but not the other way around.

Ideally it is desired that: (i) $\omega C_{gs} R_i \ll 1$ and (ii) $\omega C_{ds} R_{ds} \gg 1$ to avoid insertion losses. While the first condition is related to the intrinsic loss of the FET (ideally $R_i$ should be as small as possible), the second condition can be met by setting the point of operation of the FET in the saturation regime [52], so that the drain-source current $I_{ds} = V_{ds}/R_{ds}$ is weakly sensitive to variations in the drain-source voltage $V_{ds} = V_1$. This is equivalent to having a very large $R_{ds}$.

For sufficiently long wavelengths, it is possible to homogenize the periodic FET loading and model the system of Fig. 1 as a uniform "medium". The FET loading contributes to the capacitance per unit of length of the homogenized line. The additional capacitance per unit of length is given by $\mathbf{Y}_F/(-i\omega d)$. Thus, the transistor effect is spread uniformly along the line. Hence, the effective capacitance matrix $\mathbf{C}_{ef}$ of the homogenized system may be written as:

$$\mathbf{C}_{ef} = \begin{pmatrix} C_{11} & 0 \\ 0 & C_{22} \end{pmatrix} + \frac{\mathbf{Y}_F}{(-i\omega d)}. \tag{10}$$

whereas the effective inductance matrix is:

$$\mathbf{L}_{ef} = \begin{pmatrix} L_{11} & 0 \\ 0 & L_{22} \end{pmatrix}. \tag{11}$$

The continuous model is expected to be accurate provided the distance between the transistors is much smaller than the operation wavelength, the so-called metamaterial regime. Later, we shall confirm the validity of this approach. In the ideal case of a non-dissipative FET operated in the saturation regime ($R_i \approx 0$ and $R_{ds} \approx \infty$), the effective capacitance reduces to:

$$\mathbf{C}_{ef} \approx \begin{pmatrix} C_{11} + \dfrac{C_{ds}}{d} & \dfrac{g_m}{-i\omega d} \\ 0 & C_{22} + \dfrac{C_{gs}}{d} \end{pmatrix}. \tag{12}$$

As seen, it has exactly the same structure as the permittivity of the 3D metamaterial [compare with Eq. (3)]. There is however a key difference: here the capacitance $C_{ef,12}$ is purely imaginary, whereas the corresponding permittivity component $\varepsilon_{xz}$ in the MOSFET-MTM was assumed real-valued in Ref. [18]. In particular, different from Ref. [18], here the system response, albeit non-Hermitian, is characterized by a broken time-reversal symmetry $\left[ \mathbf{C}_{ef}(\omega) \neq \mathbf{C}_{ef}(\omega^*)^* \right]$. Note that the response is also nonreciprocal because $\mathbf{C}_{ef}$ does not have transpose symmetry.

## A. Transmission line modes

The wave propagation in the homogenized lines is described by the linear system (7) with the effective capacitive and inductive elements determined by Eqs. (10) and (11), respectively. We can write Eqs. (7a-d) in a compact form as $i\partial_y \mathbf{f} = \mathbf{M} \cdot \mathbf{f}$, with the matrix $\mathbf{M}$ given by

$$\mathbf{M} = -\omega \begin{pmatrix} 0 & 0 & L_{11} & 0 \\ 0 & 0 & 0 & L_{22} \\ C_{ef,11} & C_{ef,12} & 0 & 0 \\ C_{ef,21} & C_{ef,22} & 0 & 0 \end{pmatrix}. \tag{13}$$

The 4-component state vector is defined by $\mathbf{f} = (\mathbf{V}, \mathbf{I})^T$, with $\mathbf{V} = (V_1, V_2)^T$ and $\mathbf{I} = (I_1, I_2)^T$. Using these notations, we can obtain the travelling wave type solutions (bulk eigenmodes with a spatial variation along $y$ of the form $e^{i\beta y}$) from the eigenvalues and eigenvectors of matrix $\mathbf{M}$, similar to the MOSFET-MTM. The system supports two eigenwaves propagating along the $+y$ direction and two eigenwaves propagating in the opposite direction.

For simplicity, in the following we focus the analysis on the particular case wherein the capacitance matrix is given by Eq. (12), so that $C_{ef,21} = 0$. In such a scenario, line 2 is perfectly isolated from line 1. Consequently, the system supports a "ordinary" wave characterized by a propagation constant $\beta_o = \omega \sqrt{L_{11} C_{ef,11}}$ and eigenvector

$$\mathbf{f}_o = (\mathbf{V}_o, \mathbf{I}_o)^T = A_o \left( \sqrt{\frac{L_{11}}{C_{ef,11}}}, 0, 1, 0 \right)^T e^{i\beta_o y}. \tag{14}$$

Note that this wave is insensitive to the cross capacitive coupling: the energy propagates exclusively in line 1 (drain line). In addition, it also supports an extraordinary wave with propagation constant $\beta_e = \omega\sqrt{L_{22}C_{ef,22}}$ and eigenvector

$$\mathbf{f}_e = (\mathbf{V}_e, \mathbf{I}_e)^T = A_e \left( \sqrt{\frac{L_{22}}{C_{ef,22}}} \frac{C_{ef,12} L_{11}}{C_{ef,22} L_{22} - C_{ef,11} L_{11}}, \sqrt{\frac{L_{22}}{C_{ef,22}}}, \frac{C_{ef,12} L_{11}}{C_{ef,22} L_{22} - C_{ef,11} L_{11}}, 1 \right)^T e^{i\beta_e y}. \tag{15}$$

This mode can be understood as a wave propagating in line 2 (gate line), which is coupled to line 1 (drain line) through the unidirectional FETs. The coupling factor $C_{ef,12}$ is determined by the transconductance of the FET isolators. In the above, $A_o$ and $A_e$ are the complex amplitudes of the ordinary and extraordinary waves, respectively. Note that the modes described above propagate along the +y-direction. The modes propagating in the opposite direction are characterized by the same propagation constants $\beta_o$ and $\beta_e$.

A generic solution of the system $i\partial_y \mathbf{f} = \mathbf{M} \cdot \mathbf{f}$ is a superposition of ordinary and extraordinary waves propagating in the coupled lines. In particular, a wave propagating along the +y-direction is of the form $\mathbf{f} = \mathbf{f}_o + \mathbf{f}_e = (\mathbf{V}_o + \mathbf{V}_e, \mathbf{I}_o + \mathbf{I}_e)^T$. Taking into account that the total power flowing in the lines is given by $P(y) = \frac{1}{2}\text{Re}\{\mathbf{V} \cdot \mathbf{I}^*\}$, one readily finds that:

$$P(y) = \frac{1}{2}\text{Re}\{\mathbf{V}_o \cdot \mathbf{I}_o^*\} + \frac{1}{2}\text{Re}\{\mathbf{V}_e \cdot \mathbf{I}_e^*\} + \frac{1}{2}\text{Re}\{\mathbf{V}_e \cdot \mathbf{I}_o^* + \mathbf{V}_o \cdot \mathbf{I}_e^*\}. \tag{16}$$

Evidently, we can identify $P_o = \frac{1}{2}\text{Re}\{\mathbf{V}_o \cdot \mathbf{I}_o^*\}$ and $P_e = \frac{1}{2}\text{Re}\{\mathbf{V}_e \cdot \mathbf{I}_e^*\}$ as the contributions of the ordinary and extraordinary waves alone, while the remaining part $P_{\text{cr}} = \frac{1}{2}\text{Re}\{\mathbf{V}_e \cdot \mathbf{I}_o^* + \mathbf{V}_o \cdot \mathbf{I}_e^*\}$ is an interference term resulting from the interaction between the two waves. Since this interference term depends on the relative phase between the propagation constants of the modes, it can give rise to a power oscillation when the waves propagate simultaneously along the line. To illustrate the "power beating" characteristic we assume that the structure represented in Fig. 1 is excited at the input ($y = 0$) of the gate line (line 2) and that the drain line (line 1) is short-circuited at the input. Both lines are terminated with matched loads to avoid reflections. For a superposition of the ordinary and extraordinary modes [Eqs. (14)-(15)] it is simple to show that $V_1(y = 0) = 0$ requires that the complex amplitudes of the waves satisfy:

$$A_o = -A_e \sqrt{\frac{C_{ef,11}}{L_{11}}} \sqrt{\frac{L_{22}}{C_{ef,22}}} \frac{C_{ef,12} L_{11}}{C_{ef,22} L_{22} - C_{ef,11} L_{11}} . \tag{17}$$

We consider that the unloaded lines are identical. The lines have characteristic impedance $Z_{01} = Z_{02} = Z_0 = 50\Omega$ and effective permittivity $\varepsilon_{ef,2} = \varepsilon_{ef,1} = \varepsilon_{ef} = 1.7$. The effective permittivity represents the permittivity of an equivalent uniform medium that substitutes both the air and dielectric regions in the microstrip line [52]. Hence, the p.u.l. capacitances and inductances of the uncoupled lines are $C_{ii} = 1/(Z_0 v_p)$ and $L_{ii} = Z_0/v_p$, respectively, with $v_p = c/\sqrt{\varepsilon_{ef}}$ the phase velocity in the lines. The lines are coupled through FET isolators characterized by [52]: $C_{gs} = 0.3\text{pF}$, $C_{ds} = 0.12\text{pF}$ and $g_m = 40\text{mS}$. Moreover, we assume that the distance between adjacent transistors (spatial period) is $d = 5\text{mm}$.

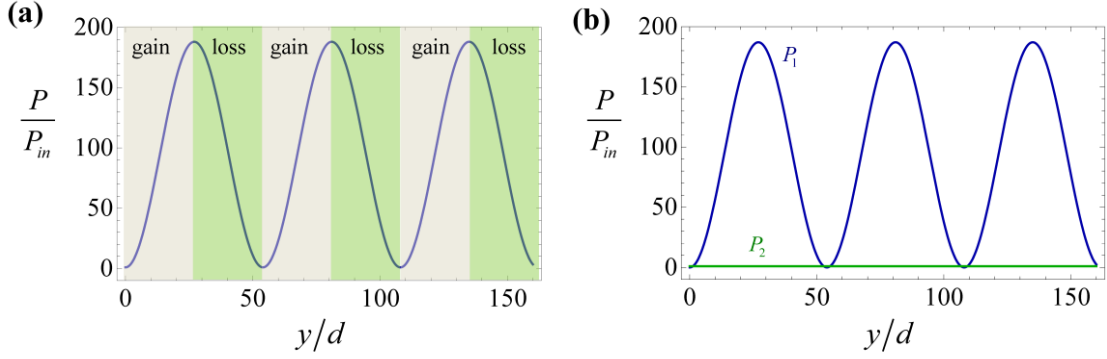

**Fig. 3. a)** Normalized power transported by the coupled lines as a function of the propagation distance normalized to the spacing $d = 5\text{mm}$ between adjacent FETs for the frequency of operation $\omega/(2\pi) = 5\text{GHz}$. The microstrip lines have the same characteristic impedance $Z_{01} = Z_{02} = Z_0 = 50\Omega$ and effective permittivity $\varepsilon_{ef,2} = \varepsilon_{ef,1} = \varepsilon_{ef} = 1.7$. The FET isolators have $R_i = 0$, $R_{ds} = \infty$, $C_{gs} = 0.3\text{pF}$, $C_{ds} = 0.12\text{pF}$, $g_m = 40\text{mS}$ and $C_{gd} = 0$. The regions shaded in green correspond to dissipative regions, whereas the light-brown shaded regions correspond to gain regions. **b)** Similar to **a)** but showing the normalized power transported by the drain line (blue curve) and gate line (green curve).

In Fig. 3a, we show the spatial distribution of the power transported in the system at the fixed frequency of operation $\omega/(2\pi) = 5\text{GHz}$. As seen, the power oscillates periodically with the propagation distance, similar to the MOSFET-MTM [18]. In the sections where the power increases (shaded brown regions in Fig. 3a) the transistors inject energy into the system yielding a substantial rise in transported power that can surpass more than 180 times the input power. Conversely, within the dissipative regions (shaded green areas in Fig. 3a), the FET isolators extract energy from the lines, resulting in a reduction of transported power. It is important to mention that despite no dissipative elements are explicitly considered in this example, the response of the FET can lead to absorption. Indeed, even in the ideal case, when $R_i = 0$ and $R_{ds} = \infty$, the transistors can extract energy from the AC signal through the current source associated with the transconductance gain, resulting in a dissipative response. Thus, different from more conventional gain systems, the MOSFET-type response can either be active or dissipative, depending on the relative phase between the voltages at the gate and drain lines.

We have also calculated the power transported by the two lines separately, given by $P_i(y) = \frac{1}{2}\text{Re}\{\mathbf{V}_i \cdot \mathbf{I}_i^*\}$, for $i = 1, 2$. The results are depicted in Fig. 3b and show that while the power transported in the gate line remains constant (green curve), the power transported in the drain line (blue curve) oscillates. This feature stems from the unidirectional properties of the transistors. The gate line is isolated from the drain line, whereas the drain line is coupled to the gate line. Thus, the power transported in the drain line results from the interference of two waves, leading to power oscillations. Similar to standard transistor-based devices, the physical origin behind the power oscillations is the energy supplied by the DC generator that biases the transistors, which can be positive or negative. The type of response, active or dissipative, depends on the relative phase of the current and voltage at the transconductance element. The transconductance element describes how the energy is transferred from the DC part of the circuit to the AC part of the circuit, through the nonlinear response of the transistor.

Up to this point, the wave propagation was described using the continuum model. It is important to consider the intrinsic granularity of the system by incorporating the periodic loading of transistors along the lines. In that case, we may study the structure as an arrangement of a finite number of unit cells. We suppose that the transistor is placed in the middle of a unit cell comprising two microstrip lines sections with length $d/2$. In that case the state vector at the input and output of the unit cell satisfies

$$\mathbf{f}(y=d) = \left( e^{-i\mathbf{M}_0 \frac{d}{2}} \cdot \mathbf{T} \cdot e^{-i\mathbf{M}_0 \frac{d}{2}} \right) \cdot \mathbf{f}(y=0), \text{ where}$$

$$\mathbf{M}_0 = -\omega \begin{pmatrix} 0 & 0 & L_{11} & 0 \\ 0 & 0 & 0 & L_{22} \\ C_{11} & 0 & 0 & 0 \\ 0 & C_{22} & 0 & 0 \end{pmatrix} \quad (18)$$

is the transfer matrix of the bare transmission lines, and

$$\mathbf{T} = \begin{pmatrix} 1 & 0 & 0 & 0 \\ 0 & 1 & 0 & 0 \\ -Y_{11} & -Y_{12} & 1 & 0 \\ -Y_{21} & -Y_{22} & 0 & 1 \end{pmatrix} \quad (19)$$

is the transfer matrix of the FET, which depends on the admittance matrix given in Eq. (9). For a finite structure with $N$ cells, one can relate the state vector at the input and output of the system as $\mathbf{f}(y = Nd) = \left( e^{-i\mathbf{M}_0 \frac{d}{2}} \cdot \mathbf{T} \cdot e^{-i\mathbf{M}_0 \frac{d}{2}} \right)^N \cdot \mathbf{f}(y = 0)$.

Next, we compare the continuum model results, with the exact solution that takes into account the discrete nature of the structure. As before, it is supposed that $V_1(y=0) = 0$. Figure 4a shows the power distribution in the lines for the same configuration as in Fig. 3. In the discrete model we only show the normalized power calculated at the input of each unit cell (consecutive discrete nodes are spaced by a distance $d$).

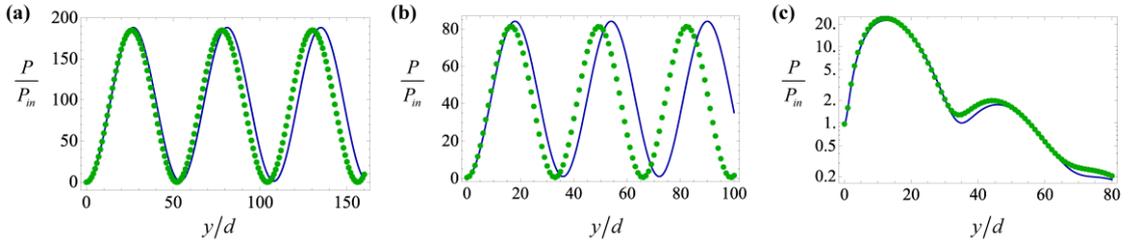

**Fig. 4**. **a)** Normalized (total) power transported by the coupled lines as a function of the propagation distance for the same coupled transmission line configuration studied in Fig. 3. **b)** Similar to **a)** but for the operation frequency $\omega/(2\pi) = 7.5\text{GHz}$. **c)** Similar to **b)** but considering the dissipative elements of the transistor model. In all plots the solid curves represent the continuous model results, and the symbols correspond to the discrete model results.

The results of Fig. 4a show a remarkable agreement between the continuous and discrete model, indicating that the continuous model describes very accurately the wave propagation in the system, particularly for short propagation distances ($y/d < 50$).

We also studied the response of the system when the frequency of operation is increased. Figure 4b depicts the results of the continuous and discrete models for a frequency of operation $\omega/(2\pi) = 7.5\text{GHz}$. As seen, compared to the configuration studied in Fig. 4a, the peak amplitude of the power transported in the lines decreases to about half, and the spatial period of the oscillations also decreases. The agreement between the two models is less good in Fig. 4b. This is an inherent limitation of the continuous model, as it is only applicable to the long wavelength regime where $\omega d/c \ll 1$. In the example of Fig. 4b, the normalized frequency is $\omega d/c \approx 0.79$.

Additionally, we investigated the impact of dissipation in the FETs on the overall response of the coupled lines. The dissipation in the FET isolators is considered by taking $R_i = 7\Omega$ and $R_{ds} = 400\Omega$ [52]. The corresponding power distribution in the lines calculated with the continuous and discrete models is shown in Fig. 4c. Interestingly, it is observed that the power in the system can experience oscillations even in the presence of the dissipative elements of the transistor model. However, the maximum gain undergoes a significant reduction, approximately fourfold, and the power beating is markedly dampened, becoming indistinct for $y/d > 80$.

## B.  *Exceptional Points*

As discussed in the previous section, the response of the coupled lines is non-Hermitian, and thereby the bulk eigenstates are non-orthogonal. Importantly, the response of a non-Hermitian system can become singular when two or more eigenstates degenerate and coalesce. This is called an exceptional point. It is easy to check that in our system such a behavior ($\mathbf{f}_e \sim \mathbf{f}_o$) emerges in the continuous model when $L_{22}C_{ef,22} = L_{11}C_{ef,11}$. In such a scenario both eigenmodes share an identical propagation constant, i.e. $\beta_o = \beta_e$, and the same field structure. At the exceptional point the general solution of $i\partial_y \mathbf{f} = \mathbf{M} \cdot \mathbf{f}$ is not generated by complex exponentials only, but rather by a combination of complex exponentials and polynomials of $y$:

$$V_1 = A_1^+ e^{i\beta_o y} + A_1^- e^{-i\beta_o y} + \frac{C_{12,ef}}{2C_{11,ef}} i\beta_o y \left[ B_2^+ e^{i\beta_o y} - B_2^- e^{-i\beta_o y} \right], \quad \text{(drain line)} \quad (20a)$$

$$V_2 = B_2^+ e^{i\beta_o y} + B_2^- e^{-i\beta_o y}, \quad \text{(gate line)} \quad (20b)$$

The currents can be found from the voltages using Eqs. (7a) and (7c). As seen, due to the term proportional to $y$, at the exceptional point the voltage in the drain line can be arbitrarily large, and hence the peak gain is not constrained by the power beating oscillation. Note that at the singular point there is a resonant interaction between the ordinary and extraordinary waves, due to the phase matching condition $\beta_o d = \beta_e d$. Specifically, $\beta_o = \beta_e$ guarantees that the relative phase between the voltage and current at the transconductance elements does not change as the wave propagates. Thus, as the wave progresses along the line, it can continuously draw power from the DC bias, different from the situation where due to $\beta_o \neq \beta_e$ some of the power already propagating in the line can be returned to the DC bias.

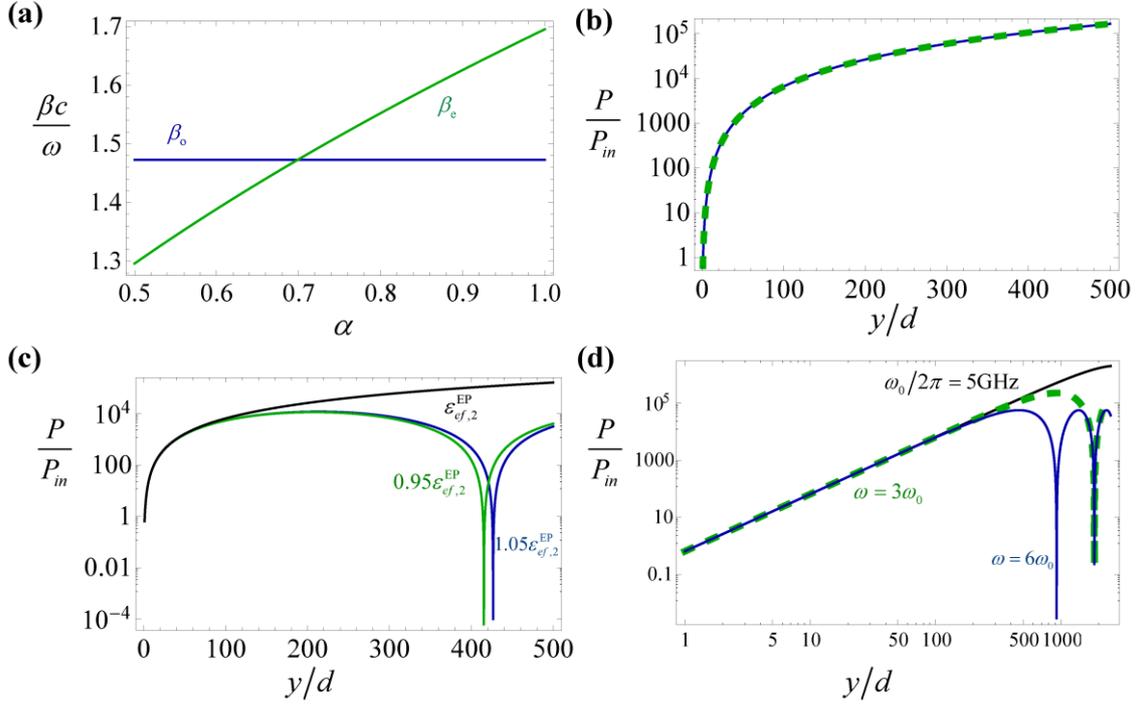

**Fig. 5**. **a)** Normalized propagation constants of the ordinary (blue curve) and extraordinary (green curve) modes calculated as a function of $\alpha = \varepsilon_{ef,2}/\varepsilon_{ef,1}$ for the same structural parameters as in Fig. 3 except for the effective permittivity of line 2 ($\varepsilon_{ef,2}$). **b)** Power gain $P/P_{in}$, when the system is operated at the exceptional point (i.e., for $\varepsilon_{ef,2}^{EP} = 0.699\varepsilon_{ef,1}$), for the structural parameters considered in **a)**. The solid blue line represents our theoretical (continuous model) results and the dashed line corresponds to the results predicted by Eq. (21). **c)** Power gain of the system under the same conditions as in **b)** but calculated using $\varepsilon_{ef,2} = \varepsilon_{ef,2}^{EP}$ (black curve), $\varepsilon_{ef,2} = 1.05\varepsilon_{ef,2}^{EP}$ (blue curve) and $\varepsilon_{ef,2} = 0.95\varepsilon_{ef,2}^{EP}$ (green curve). **d)** Power gain of the system under the same conditions as in **b)** but calculated at different frequencies $\omega/(2\pi) = \omega_0/(2\pi) = 5\text{GHz}$ (black solid curve), $\omega = 3\omega_0$ (green dashed curve) and $\omega = 6\omega_0$ (blue solid curve) and considering $C_{gd} = 0.01\text{pF}$.

In order to investigate the properties of the exceptional point in our system we consider the same parameters and operation frequency as in Fig. 3, except that now we suppose that input line (line 2) is characterized by an effective permittivity $\varepsilon_{ef,2} = \alpha\varepsilon_{ef,1}$. As before, the line impedances are $Z_{01} = Z_{02} = Z_0 = 50\Omega$. In Fig. 5a, we depict the propagation constants of the ordinary and extraordinary waves as a function of $\alpha$. As seen, an exceptional point emerges for $\alpha = 0.699$ ($\varepsilon_{ef,2}^{\text{EP}} = 1.1883$), which ensures that both the two propagation constants and the two eigenmodes coalesce.

At the exceptional point, the Bloch-wave type description breaks down, as Eq. (15) becomes singular. The modes at the exceptional point have a polynomial nature, as shown in Eq. (20). Hence, by operating the system at an exceptional point the power drawn from the DC supply is maximized. As the spacing between adjacent FETs is the same on both lines, the exceptional point $\beta_o = \beta_e$ occurs when the phase delay in the lines is identical. This is often referred to as synchronization condition in the design of microwave amplifiers, such as the distributed amplifier [52]. From Eq. (20a) it is clear that when the lines are terminated with matched loads and for sufficiently large $y$ we can take $V_1 \approx \dfrac{C_{12,ef}}{2C_{11,ef}} i\beta_o y B_2^+ e^{i\beta_o y} = \dfrac{C_{12,ef}}{2C_{11,ef}} i\beta_o y V_2$. Thus, the output (line 1) and input (line 2) powers are related as $P_{out}/P_{in} \approx \left|\dfrac{C_{12,ef}}{2C_{11,ef}}\right|^2 |\beta_o y|^2 \dfrac{Z_g}{Z_d} = g_m^2 \dfrac{Z_g Z_d}{4}\left(\dfrac{y}{d}\right)^2$ with $Z_g = \sqrt{L_{22}/C_{ef,22}}$, $Z_d = \sqrt{L_{11}/C_{ef,11}}$. The power flowing in the line grows quadratically with the length of the line. The derived result can also be written as:

$$G = \frac{P_{out}}{P_{in}} = \frac{g_m^2 Z_d Z_g N^2}{4}. \tag{21}$$

where $N$ is the number of FETs in the structure. Curiously, the above formula is well known in the design of distributed amplifiers [52], which thereby are operated at an exceptional point. The power gain grows quadratically with the number of FETs, and also depends on the transconductance gain and on the characteristic impedance of the effective lines. It is relevant to mention that in most practical scenarios the synchronization condition is enforced by varying the distance between successive FET, through the meandering the transmission lines. Here, we consider lines with different dielectric substrates merely to simplify the analytical modeling.

In Fig. 5b we represent the power characteristic of the system at the exceptional point calculated with the continuous model. As seen, the results predicted by Eq. (21) overlap precisely the effective transmission line results. Importantly, even the slightest detuning from the operation at the exceptional point will break the phase synchronization condition and reveal a power beating characteristic. This is shown in Fig. 5c, where we depict the power gain of the system as a function of the number of transistors when the effective permittivity of line 2 is changed from $\varepsilon_{ef,2} = \varepsilon_{ef,2}^{EP}$ (black curve), to $\varepsilon_{ef,2} = 1.05\varepsilon_{ef,2}^{EP}$ (blue curve) and $\varepsilon_{ef,2} = 0.95\varepsilon_{ef,2}^{EP}$ (green curve). As seen, when the system is not operated at the exceptional point, the oscillations reappear, as the interaction between the ordinary and extraordinary waves is not strictly "constructive" for all points of space. Nevertheless, for all practical purposes such effect would go unnoticed as typical distributed amplifiers do not have more than a dozen transistors [53-58]. The robustness of the power gain for small systems is precisely the reason why such amplifiers are considered to be wideband [49-58].

In Fig. 5d, we depict the calculated power gain when the capacitance $C_{gd}$ is not neglected (we use $C_{gd} = 0.01\text{pF}$ [52]) and the frequency of operation is changed from $\omega_0/(2\pi) = 5\text{GHz}$ (black solid curve) to $\omega = 3\omega_0$ (green dashed curve) and $\omega = 6\omega_0$ (blue solid curve). As seen, even when the FETs do not behave as ideal isolators and the system is no longer operated at the exceptional point, for structures with $N_{FET} = 100$ the system response remains nearly the same as in the ideal scenario, despite the variations in the operation frequency. Indeed, in most practical implementations the power gain is only limited by dissipation [52] and by the frequency response of the transistors, which for microwave FETs tends to breakdown around 20GHz [59].

## IV. Conclusions

We proposed a 1D implementation of the nonreciprocal and non-Hermitian MOSFET-MTM based on transmission lines capacitively coupled through FET isolators. We studied the wave propagation in the system under an effective medium approach and demonstrated that the homogenized system response is non-Hermitian and characterized by a broken time-reversal symmetry. It was shown that the interaction between the eigenmodes in the structure can lead to regimes wherein the coupled lines enable gain and other regimes where they act as lossy. We also demonstrated that due to the non-Hermitian response, the system parameters can be tuned in such a way that the structure is operated at an exceptional point, wherein both eigenstates coalesce and degenerate. At the exceptional point the wave propagation is not described by "Bloch waves", but rather by polynomial-type Bloch waves. In this regime, there are no power oscillations and the wave can continuously draw power from the DC bias. The voltage and the current grow linearly with the propagation distance, while the power grows quadratically. Furthermore, we demonstrated that typical distributed amplifiers are operated exactly at the exceptional point.

**Acknowledgements**

This work was partially funded by the Institution of Engineering and Technology (IET) under the A F Harvey Research Prize 2018, by the Simons Foundation, and by Instituto de Telecomunicações under Project No. UID/EEA/50008/2020. D. E. Fernandes acknowledges financial support by IT-Lisbon under the research contract with reference C-0042-22. S. Lannebère acknowledges financial support by IT-Coimbra under the research contract with Reference DL 57/2016/CP1353/CT0001. T. A. Morgado acknowledges FCT for research financial support with reference CEECIND/04530/2017 under the CEEC